\documentclass[11pt,onecolumn,amssymb,nofootinbib]{revtex4}
\usepackage{amsmath, amsthm, amscd, amssymb}
\usepackage{bm}
\usepackage{bbm}

\begin{document}

\title{\bf Can the flyby anomaly be attributed to earth-bound dark
matter?}
\author{Stephen L. Adler}
\email{adler@ias.edu} \affiliation{Institute for Advanced Study,
Einstein Drive, Princeton, NJ 08540, USA.}

\begin{abstract}
We make preliminary estimates to assess whether the recently
reported flyby anomaly can be attributed to dark matter
interactions. We consider both elastic and exothermic inelastic
scattering from dark matter constituents; for isotropic dark
matter velocity distributions,  the former  decrease, while the
latter increase, the final flyby velocity.  The fact that  the observed
flyby velocity anomaly shows examples with both positive and
negative signs, requires the dominance of different dark matter scattering processes
along different flyby trajectories.
The magnitude of the observed anomalies requires
dark matter densities many orders of magnitude greater than the
galactic halo density. Such a large density could result from an
accumulation cascade, in which the solar system-bound dark matter
density is much higher than the galactic halo density, and the
earth-bound density is much higher than the solar system-bound
density.  We discuss a number of strong constraints on the hypothesis of a dark matter
explanation for the flyby anomaly. These require dark matter to be non-self-annihilating,
with the dark matter scattering cross section on
nucleons much larger, and the dark matter mass much lighter, than usually assumed.

\end{abstract}
\maketitle

\subsection{Introduction}

In a recent paper, Anderson et al. \cite{and} have reported
anomalous orbital energy changes, of order 1 part in $10^6$,
during earth flybys of various spacecraft. Some flybys  show
energy decreases, and others energy increases, with the signs and
magnitudes related to the spacecraft initial and final velocity
orientation with respect to the equatorial plane. Since the
DAMA/LIBRA collaboration \cite{DAMA} has recently reported an annual modulation
signal interpreted as
evidence for galactic halo dark matter, it is natural to ask
whether the flyby anomalies could be attributed to dark matter
interactions.  In this paper we give some preliminary calculations
directed at this question. Needless to say, in proceeding along
this route we are assuming that the reported flyby anomalies are
not artifacts of the orbital fitting method used in \cite{and}.  For
a detailed discussion of this, and further references, see \cite{lam},
which concludes that the most obvious candidates for artifactual
explanations cannot give the large effect observed.   This of course
does not rule out the possibility that something has been overlooked, and searching
for a conventional explanation of the flyby anomaly
is clearly a line of investigation that should be vigorously pursued.\footnote{One possibility being discussed, and raised by a referee of this paper,  is that the reported anomaly may arise from a mismodeling  of the
earth's reference frame within the barycentric system, since the earth's position
relative to the sun is not known to a precision better than a kilometer.  While it
will be important to test the effects of this  imprecision on integrations of
the flyby trajectory, an argument based on energy conservation suggests that
it will be too small.  The magnitude of the change in the flyby potential energy per unit mass  in the sun's
gravitational field is $\sim GM_{\odot} \Delta R/A^2$, with $\Delta R \sim
1.4 \times 10^5 {\rm km}$ the distance travelled by the flyby between ingoing and
outgoing asymptotes, with $A \sim 1.5 \times 10^8 {\rm km}$ the earth-sun distance,
and with $GM_{\odot} \sim 1.3 \times 10^{11} {\rm km}^3 {\rm s}^{-2}$.   The error in this
potential energy change arising from an uncertainty $\delta A \sim 1 {\rm km}$ in $A$ is then $\sim 2GM_{\odot} \Delta R \delta A/A^3 \sim  10^{-8} {\rm km}^2
{\rm s}^{-2}$.  However, the magnitude of the anomaly in the flyby kinetic energy
per unit mass is $\vec v_f \cdot \delta \vec v_f \sim 10^{-6} (10\, {\rm km}\,{\rm s}^{-1})^2 \sim 10^{-4} {\rm km}^2 {\rm s}^{-2}$, which is four orders of magnitude
larger than the error in the sun's potential energy arising from the uncertainty
in the earth's position.  A similar calculation for the moon shows that the
$\sim 2 {\rm cm}$ uncertainty in its position relative to the earth makes a contribution eight orders of magnitude smaller than the flyby anomaly.}

\subsection{Elastic and Inelastic Dark Matter Scattering}

Let us consider the velocity change when a spacecraft
nucleon\footnote{ Dark matter scattering from electrons would also
be expected, with subsequent sharing of the momentum change with
nucleons, but since this is harder to model, we ignore it for the
purpose of making order of magnitude estimates.} of mass $m_1\simeq
1 {\rm GeV}$ and initial velocity $\vec u_1$ scatters from a primary
dark matter particle of mass $m_2$ and initial velocity $\vec u_2$,
into an outgoing nucleon of mass $m_1$ and velocity $\vec v_1$, and
an outgoing secondary dark matter particle of mass $m_2'=m_2-\Delta
m$ and velocity $\vec v_2$ . The inelastic case corresponds to
$m_2'\not= m_2$, while in the elastic case, $m_2'=m_2$ and $\Delta
m=0$. (The possible relevance of inelastic scattering has been
emphasized in a recent paper of Bernabei et al. \cite{inel}; see
also the book of Khlopov \cite{khlop1}, which gives arguments for
unstable dark matter particles and reviews proposals \cite{khlop2}
that dark matter may consist of ``mirror'' particles.) Under the
assumptions, (i) both initial particles are nonrelativistic, so that
$|\vec u_1|<<c, |\vec u_2|<<c$, and (ii) the center of mass
scattering amplitude $f(\theta)$ depends only on the auxiliary polar
angle $\theta$ of scattering\footnote{Here $\theta$ is the
kinematically free angle between $\vec v_1-(m_1\vec u_1+m_2 \vec
u_2)/(m_1+m_2')$ and $\vec u_1-(m_1\vec u_1+m_2\vec u_2)/(m_1+m_2)$,
which when $m_2' \neq m_2$ is not the same as the angle between the
incident and outgoing nucleon in the center of mass frame.}, a
straightforward calculation shows that the outgoing nucleon velocity
change, averaged over scattering angles, is given by
\begin{equation}\label{eq:vel}
\langle \delta \vec v_1\rangle =\frac {m_2 \vec u_2 - m_2' \vec
u_1}{m_1+m_2'}+t\langle \cos \theta \rangle \frac {\vec u_1-\vec
u_2}{|\vec u_1-\vec u_2|}~~~,
\end{equation}
with $t>0$ given by taking the square root of
\begin{equation}\label{eq:tdef}
t^2=\frac{m_2m_2'}{(m_1+m_2)(m_1+m_2')}(\vec u_1-\vec u_2)^2 +
\frac{\Delta m~ m_2'}{m_1(m_1+m_2')} \Big[2 c^2 - \frac{(m_1\vec
u_1+m_2\vec u_2)^2}{(m_1+m_2)(m_1+m_2')}\Big]~~~,
\end{equation}
and with $\langle \cos \theta \rangle$ given by
\begin{equation}\label{eq:costhetdef}
 \langle \cos \theta \rangle = \frac {\int_0^{\pi}d\theta \sin
\theta \cos \theta |f(\theta)|^2} {\int_0^{\pi}d\theta \sin \theta
|f(\theta)|^2}~~~.
\end{equation}
In the elastic scattering case, with $\Delta m=0$, $m_2'=m_2$,
these equations simplify to
\begin{equation}\label{eq:el}
\langle \delta \vec v_1\rangle = -2\frac{m_2}{m_1+m_2}(\vec
u_1-\vec u_2)\langle \sin^2(\theta/2) \rangle~~~.
\end{equation}
In the inelastic case, assuming that  $\Delta m/m_2$ and
$m_2'/m_2$ are both of order unity,  the equations are well
approximated by
\begin{equation}\label{eq:inel}
\langle \delta \vec v_1\rangle \simeq \frac{\vec u_1-\vec
u_2}{|\vec u_1-\vec u_2|}     \Bigg( \frac{2 \Delta m ~m_2'} {
 m_1 (m_1+m_2')}\Bigg)^{1/2}c \langle \cos \theta \rangle ~~~.
\end{equation}
Since $\vec u_1$ and $\vec u_2$ are typically of order 10 ${\rm
km} ~{\rm s}^{-1}$, the velocity change in the inelastic case is
larger than that in the elastic case by a factor $ \sim c/|\vec
u_1|\sim 10^4.$

To get the force per unit spacecraft mass resulting from dark
matter scatters, that, is the acceleration, one multiplies the
velocity change in a single scatter $\langle \delta \vec
v_1\rangle $ by the number of scatters per unit time.  This latter
is given by the flux $|\vec u_1-\vec u_2|$, times the scattering
cross section $\sigma$, times the dark matter spatial and velocity
distribution $\rho\big(\vec x, \vec u_2\big)$.  Integrating out
the dark matter velocity, one thus gets for the force acting at
the  point $\vec x(t)$ on the spacecraft trajectory with velocity
$\vec u_1=d\vec x(t)/dt$,
\begin{equation}\label{eq:force}
\delta \vec F= \int d^3 u_2 \langle \delta \vec v_1\rangle |\vec
u_1-\vec u_2| \sigma \rho\big(\vec x, \vec u_2\big)~~~.
\end{equation}
Equating the work per unit spacecraft mass along a trajectory from
$t_i$ to $t_f$ to the change in kinetic energy per unit mass
(assuming that the initial and final times are in the asymptotic
region where the potential energy can be neglected) we get
\begin{align}\label{eq:work}
\delta \frac{1}{2}(\vec v_f^{\,2} -\vec v_i^{\,2}) = &\vec v_f
\cdot \delta \vec v_f = \int_{t_i}^{t_f} dt  (d\vec x/dt) \cdot
\delta \vec F\cr =&\int_{t_i}^{t_f} dt \int d^3 u_2 (d\vec x/dt)
\cdot \langle \delta \vec v_1\rangle |\vec u_1-\vec u_2| \sigma
\rho\big(\vec x, \vec u_2\big)~~~.\cr
\end{align}
To get the vectorial change in velocity is more difficult; one
must solve the perturbed orbital differential equation (taking
here the center of the earth as the origin of coordinates),
\begin{equation}\label{orbit}
 \frac {d^2 \delta \vec x}{(dt)^2}= -\frac {G M_{\oplus} }{|\vec
 x|^3} \Bigg(\delta \vec x - \frac{3 \vec x\cdot \delta \vec x
 \, \vec x}{|\vec x|^2}\Bigg) + \delta \vec F~~~.
 \end{equation}
 One can check that taking the inner product of this equation with
 $d\vec x/dt$, integrating over time, and integrating by parts twice,
  again gives the energy
 conservation relation $\vec v_f \cdot \delta \vec v_f =
 \int_{t_i}^{t_f} dt d\vec x/dt \cdot \delta \vec F$.

Consider now the case of a dark matter density that has an
inversion invariant velocity distribution, so that $ \rho\big(\vec
x,\vec u_2\big)=\rho\big(\vec x,-\vec u_2\big)$. From
\eqref{eq:el} and \eqref{eq:work}, we see that in the elastic
case, the flux weighting factor favors $\vec u_2$ being oppositely
directed to $\vec u_1=d\vec x(t)/dt$, and so the flyby velocity change, integrated over the dark matter velocity distribution,  is oppositely directed to $d\vec x(t)/dt$. Hence, as
expected for elastic scattering, one gets a positive drag
coefficient and the net effect is a reduction in spacecraft
velocity. Turning to the inelastic case, where the flux factor in
\eqref{eq:work}  cancels the denominator $|\vec u_1-\vec u_2|$ in
\eqref{eq:inel}, the integration over $\vec u_2$ leaves only the
term $\vec u_1=d\vec x(t)/dt$, and so in this case the
 flyby velocity change, integrated over the dark matter velocity distribution,  is parallel to $d\vec x(t)/dt$
when $\langle\cos \theta \rangle>0$.
So for forward dominated exothermic inelastic scattering, the drag
coefficient is negative and the net effect is an increase in
spacecraft velocity, while for backward dominated inelastic scattering,
the drag coefficient is positive, as in the elastic case.  Since the observations reported in \cite{and}
show cases of increased velocity, and of decreased velocity,  a
dark matter explanation (assuming an approximately
isotropic velocity distribution) requires the presence, in differing
proportions on different trajectories, of inelastic forward dominated
scattering, and of either elastic or inelastic backward dominated
scattering.  This could be achieved in a two-component
dark matter model, with differing spatial
densities $\rho(\vec x,\vec u_2)$ governing the inelastic and
elastic scatterers.  Another possibility is a single dark matter component with an anisotropic velocity distribution, undergoing
inelastic scattering, and possibly also elastic scattering as well.
Detailed modelling will be needed to see which possibilities are viable.\footnote{In the inelastic case, the early universe   populations of the dark matter primary of mass $m_2$ and the dark matter secondary of mass $m_2^{\prime}$ will be in an equilibrium resulting from dark matter scattering from nucleons or quarks.  Hence it is not unreasonable to assume that
populations of both types of particles could survive to the present epoch, as would be needed for a two-component model.}

\subsection{Quantitative estimates}

Let us now turn to some quantitative estimates. To get a velocity
change of order $10^{-6}$ of the spacecraft velocity over a time
interval $T$ one needs
\begin{equation}\label{eq:totch}
10^{-6} \sim T \bar f \sigma \bar \rho |\langle \delta \vec v_1
\rangle| /|\vec v_f|~~~,
\end{equation}
with $\bar f$ the average flux, $\bar \rho$ the average dark
matter density, $\sigma$ the scattering cross section, and
$|\langle \delta \vec v_1 \rangle|$ the magnitude of the single
scattering velocity changes given, in the elastic and inelastic
cases, by \eqref{eq:el} and \eqref{eq:inel} respectively. This
gives an estimate of the required product of mean dark matter
density times interaction cross section,
\begin{equation}\label{eq:totch1}
\sigma \bar \rho \sim 10^{-6}|\vec v_f| /( T \bar f   |\langle
\delta \vec v_1 \rangle|)~~~.
\end{equation}.

Anderson et al. \cite{and} report that for the NEAR spacecraft
flyby, the velocity change occurs during an interval $T=3.7 {\rm
h}\sim 10^4 {\rm s}$ when the spacecraft could not be tracked
during near earth approach. Taking this estimate for $T$ and
taking the mean flux as $\bar f \sim 10 {\rm km}~{\rm s}^{-1} =
10^6 {\rm cm}~{\rm s}^{-1}$, \eqref{eq:totch1} becomes
\begin{equation}\label{eq:totch2}
\sigma \bar \rho \sim 10^{-16} {\rm cm}^{-1} |\vec v_f|/ |\langle
\delta \vec v_1 \rangle|~~~.
\end{equation}
Defining the mean dark matter mass density as $\bar \rho_m = m_2
\bar \rho$, using \eqref{eq:el} gives for the elastic case
\begin{equation}\label{eq:eldens}
\sigma \bar \rho_m \sim 10^{-16} {\rm cm}^{-1}(m_1+m_2) \geq
10^{-16}({\rm GeV}/c^2) {\rm cm}^{-1}~~~,
\end{equation}
while using \eqref{eq:inel} gives for the inelastic case\footnote{For an inelastic
exothermic reaction, the cross section increases as $1/v$ for small incident
velocities $v$ \cite{wein1}, and thus the product $\sigma v$ is what is well-defined
near threshold.  Rewriting \eqref{eq:ineldens} in terms of $\sigma \bar f$, with
$\bar f \sim 10 {\rm km}\, {\rm s}^{-1}$ the flux, we have
$\sigma \bar f \rho_m \geq 10^{-14} ({\rm GeV}/c^2) \, {\rm s}^{-1}$.}
\begin{equation}\label{eq:ineldens}
\sigma (\Delta m m_2')^{1/2} \bar \rho \sim \sigma \bar \rho_m
\sim 10^{-20} {\rm cm}^{-1} [m_1(m_1+m_2')]^{1/2} \geq 10^{-20}
({\rm GeV}/c^2) {\rm cm}^{-1}~~~.
\end{equation}

To estimate dark matter densities from these bounds, we must
assume a value for the scattering cross section.  For a cross
section of order 1 picobarn =$10^{-36} {\rm cm}^2$, we get dark
matter mass densities $\bar \rho_m \sim 10^{20} ({\rm GeV}/c^2)\,{\rm
cm}^{-3}$ in the elastic case, and  $\bar \rho_m \sim 10^{16} ({\rm
GeV}/c^2)\,{\rm cm}^{-3}$ in the inelastic case.  For a
cross section of order 1 millibarn  =$10^{-27} {\rm cm}^2$, which
would require dark matter masses much below a GeV, we get
corresponding dark matter mass densities $\bar \rho_m \sim 10^{11}
({\rm GeV}/c^2)\,{\rm cm}^{-3}$ in the elastic case, and $\bar \rho_m
\sim 10^{7} ({\rm GeV}/c^2)\,{\rm cm}^{-3}$ in the inelastic case.
These dark matter mass density bounds  are orders of magnitudes
larger than the estimated galactic halo dark matter mass density
of $0.3 ({\rm GeV}/c^2) {\rm cm}^{-3}$, but are still many orders of
magnitude smaller than the earth mass density of about $3 \times 10^{24}
({\rm GeV}/c^2){\rm cm}^{-3}$.

Can such large dark matter densities exist in orbit around the earth?
In a separate note \cite{dark}, we have pointed out that by comparing
the total mass (in gravitational units) of the earth-moon system, as
determined by lunar laser ranging, with the sum of the lunar mass as independently
determined by its gravitational action on satellites or asteroids, and the earth
mass as determined by the LAGEOS geodetic survey satellite, one can get
a direct measure of the mass of earth-bound dark matter lying between the radius of
the moon's orbit and the geodetic satellite orbit.  Current data show that the
mass of such earth-bound dark matter must be less than $4 \times 10^{-9}$ of
the earth's mass, giving an upper dark matter mass limit of $1.3 \times 10^{43}
{\rm GeV}/c^2$.
 To explain the flyby anomalies, earth-bound dark
matter would have to be concentrated within a radius of about
70,000 km around earth, which contains a volume of $\simeq 1.4 \times 10^{30}
{\rm cm}^3$; for the dark matter mass within this
volume not to exceed $4 \times 10^{-9} M_{\oplus}$, the mean dark matter density would have
to be bounded by about $  10^{13} ({\rm GeV}/c^2) {\rm cm}^{-3}$. By the above
estimates, this would correspond, in the inelastic case, to a cross section
$\sigma> 10^{-33} {\rm cm}^2$, and in the elastic case, to a cross section
$\sigma> 10^{-29} {\rm cm}^2$.  These cross sections are much larger than usually
assumed for the interactions of dark matter with nucleons, but can be compatible
with existing bounds on dark matter interaction cross sections if the dark
matter mass is much below a GeV.

\subsection{Accumulation cascade}

Because  earth-bound dark matter mass densities of order $\bar
\rho_m \sim 10^{7} ({\rm GeV}/c^2){\rm cm}^{-3}$ or larger greatly
exceed the estimated galactic halo dark matter mass density, a
mechanism for concentrating dark matter near earth would be
needed. One possibility is an accumulation cascade, in which solar
system-bound dark matter is accumulated over the lifetime of the
solar system, and then this enhanced dark matter density leads to
a further accumulation near earth. Bearing in mind that it is
an open question whether there are
efficient mechanisms for dark matter capture by the solar system
or earth  \cite{kr}, \cite{ann}, \cite{xu}  we nonetheless proceed
to estimate whether such a mechanism, with a high capture
fraction, could lead to the dark matter densities needed to
explain the flyby discrepancies.  We note also that Fr\`ere, Ling, and
Vertongen \cite{frere} have pointed out that local dark matter concentrations in
the galaxy may have played a role in the formation of the solar system, which could
give another mechanism for producing a higher sun-bound or earth-bound dark matter density than
the mean galactic halo density.

Let us start with the solar system, which is moving through the
galaxy at a velocity of $v_{\rm s.s.}\sim 220 {\rm km}~{\rm
s}^{-1}$, with the local galactic halo dark matter approximated by
a Maxwellian velocity distribution with a r.m.s. velocity of
similar magnitude. Let $f_{\rm s.s.}$ be the probability of
capture of a dark matter particle near a solar system earth orbit
of radius $A\equiv 1 {\rm a.u.} \simeq 1.5 \times 10^{8} {\rm
km}$. Then assuming particles captured in an annulus of radius $A$
and area $2 \pi A dA$ over the solar system lifetime $T_{\rm s.s.}
\sim 1.5 \times 10^{17} {\rm s}$  are redistributed, over time,
into a volume $4 \pi A^2 dA$, the captured particle mass density
at radius $A$ would be
\begin{equation}\label{eq:sscapture}
\rho_{m; {\rm s.s.}}/  \rho_{m; {\rm halo}} \sim \frac {f_{\rm
s.s.}}{2A} v_{\rm s.s.} T_{\rm s.s.} \sim 10^{11}f_{\rm s.s.} ~~~.
\end{equation}
So for $f_{\rm s.s.}$ of unity, a very large concentration of dark
matter particles in the solar system would be possible.  In fact,
the known limits on a local excess of solar system dark matter \cite{frere}, \cite{khrip} are
about $3 \times 10^5$ times the galactic halo mass density, so
$f_{\rm s.s.}$ in \eqref{eq:sscapture} could be of order $10^{-5}$
at most. We remark in passing that an enhanced solar system
density of dark matter particles would show up as a daily sidereal
time modulation of dark matter particle counting rates in
sufficiently sensitive experiments of the DAMA/LIBRA type, just as
the galactic halo dark matter density is detected  by DAMA/LIBRA
as an annual modulation \cite{annmod} in the counting rate.

Given an enhanced solar system dark matter density, we can now
make a similar estimate of the maximum possible capture density in
an earth orbit, by replacing $f_{\rm s.s}$ by the corresponding
earth capture fraction $f_{\rm e}$, replacing $v_{\rm s.s.}$ by
the orbital velocity of earth around the sun $v_{\rm e} \sim 30
{\rm km}~{\rm s}^{-1}$, and replacing $A$ by the  earth orbit
radius relevant for the flyby anomalies, $R \sim 7 \times 10^4
{\rm km}$. This gives
\begin{equation}\label{eq:eacapture}
\rho_{m; {\rm e}}/  \rho_{m; {\rm s.s.}} \sim \frac {f_{\rm
e}}{4R} v_{\rm e} T_{\rm s.s.} \sim 2 \times 10^{13} f_{\rm e}~~~,
\end{equation}
where we have divided by an extra factor of 2 since we are
assuming that the solar system dark matter density is linearly
increasing over its lifetime. So if the solar system dark matter
density were equal to its upper bound,  and $f_{\rm e}$ were of
order unity, the earth-bound dark matter density at or below the
radius relevant for the flyby anomalies could be as large as $\sim
10^{19}$ times the galactic halo density. So even with small
values of $f_{\rm e}$, one could attain large enough values of
dark matter density to explain the flyby discrepancy if the
interaction cross section were large enough.

\subsection{Constraints}

In addition to having to provide a large enough dark matter density,
such a mechanism would have to lead to a dark matter spatial
distribution satisfying significant  constraints.  We shall consider
three types of constraints, (1) constraints coming from data on closed orbits of
satellites, the moon, and the earth, (2) constraints coming from stellar
dynamics,  and (3) constraints coming from earth and satellite heating.

\subsubsection{Closed orbit constraints}

We begin with an analysis of closed orbit constraints, by asking  what is the most general form of a drag force that gives zero
cumulative drag for all closed satellite orbits.  Let us rewrite \eqref{eq:work}
for the work per unit spacecraft mass as
\begin{equation}\label{eq:work1}
\delta W= \int dt (d\vec x/dt) \cdot \delta \vec F =
\int d\theta (d\vec x/d\theta) \cdot \delta \vec F~~~,
\end{equation}
with $\theta$ the angle in the orbital plane between the orbit semi-major axis and the vector from the earth's center to the  satellite, and let us define the ``drag function'' $D(\vec x,\vec v=d\vec x/dt)$ as
\begin{equation}\label{eq:dragfn}
D(\vec x,\vec v)=(d\vec x/d\theta) \cdot \delta \vec F~~~.
\end{equation}
Then the  condition for vanishing cumulative drag over the orbit  becomes
\begin{equation}\label{eq:work2}
\int_0^{2\pi} d\theta D\big(\vec x(\theta),\vec v(\theta)\big)=0~~~.
\end{equation}
 Since each pair $\vec x,\vec v$ is Cauchy data that corresponds to a distinct orbit, a general solution
to \eqref{eq:work2} is
\begin{equation}\label{eq:gensoln}
D(\vec x,\vec v)=\sum_{\ell=1}^{\infty}(a_{\ell}\sin\ell \theta +b_{\ell} \cos\ell\theta)~~~,
\end{equation}
with $\theta$ determined by $\vec x,\vec v$ and with the coefficients $a_{\ell},\,
b_{\ell}$ functions of the five orbit constants of motion (angular momentum vector, energy, and semi-major axis orientation) that in turn can be computed as functions
of $\vec x,\,\vec v$.\footnote{This statement applies to $\vec x,\vec v$ values that do not correspond to earth-intersecting orbits.}   That is, \eqref{eq:work2} is satisfied by requiring that
the Fourier series expansion in $\theta$ of the drag function has no constant term $b_0$.   For a
hyperbolic orbit such as the flyby orbits, the cumulative energy change per unit
spacecraft mass is obtained by integrating \eqref{eq:work1} from $-\theta_D$ to $\theta_D$, with $2\theta_D$ the flyby deflection angle, giving
\begin{equation}\label{eq:work3}
\delta\frac{1}{2}(\vec v_f^{\,2}-\vec v_i^{\,2})=2 b_0 \theta_D +
2\sum_{\ell=1}^{\infty} \frac {b_{\ell}}{\ell} \sin \ell \theta_D~~~,
\end{equation}
where we have included the possibility of a nonzero $b_0$.
Details of how the near-earth environment (such as a hypothetical dark matter
distribution) influence the flyby energy change appear through the coefficient
functions $b_{\ell}$.  In particular, the fitting formula given in \cite{and}
would have to arise this way, through the dynamics determining  the coefficients $b_{\ell}$, and not through the kinematics of requiring vanishing drag anomaly for closed orbits,
corresponding to vanishing $b_0$.\footnote{For example, one might try a dynamical
model in which there are inelastic and elastic scatterers with roughly similar
values of scattering cross section times density, with the inelastic
scatterers distributed in a prolate ellipsoid,  elongated towards
the poles, and the elastic scatterers in a oblate ellipsoid,
somewhat elongated towards the equator. For circular satellite
orbits in the overlap region of the two distributions, the positive
and negative drag effects would cancel; for a flyby deflected from a
small to a large angle with respect to the equator, the negative
drag effects would predominate, giving a velocity increase. For a
flyby deflected from a moderate to  a smaller angle with respect to
the equator, the positive drag effects would dominate, giving a
velocity decrease.  Such a model would not reproduce the fitting
formula given in \cite{and}, but with appropriate shapes of the
density profiles might be able to accommodate the six flyby data
sets used to generate that fit.  Clearly, this is but one example of many
possible scenarios.}

This analysis suggests that if the flyby effect is confirmed, there likely will be
analogous drag anomalies in high-lying satellite orbits, since a vanishing $b_0$
would require a ``fine-tuning'' in the drag law, with cancelling negative and positive drag contributions around closed orbits.
Since normal satellite atmospheric drag
effects are proportional to the cross-sectional area of the satellite,
whereas dark matter scattering drag (of either sign) is proportional
to the mass of the satellite, it would be helpful to have an analysis
of drag effects in existing satellites, assuming the presence of both area-proportional and mass-proportional components.  The aim would be  to see if there is any
evidence for small mass-proportional drag contributions, or at least to place
bounds on such contributions for use as constraints on dark matter model fits to the flyby data.

Ignoring now the possibility of fine-tuning of the drag force that could give cancellation between negative and positive
contributions over
closed orbits, let us analyze several
constraints that come from observation of the rate at which the radius of an orbiting body increases or decreases, which can be used to bound a drag force acting on it as follows.  For definiteness let us consider the earth's orbit around the sun,
since other cases can be obtained from this by appropriate substitutions.
Approximating the earth's orbit as circular, the total energy (potential plus kinetic) is $E=-GM_{\oplus}M_{\odot}/(2A)$, and the orbital velocity
is given by $v_e^2=GM_{\odot}/A$, from which one easily derives that
\begin{equation}\label{eq:earthen}
\frac {dE}{M_{\oplus}c^2}=\frac{1}{2} \frac{v_e^2}{c^2} \frac{dA}{A}~~~.
\end{equation}
Letting $dA$ be the change in $A$ over a single orbit, the left hand side
of \eqref{eq:earthen} is given by $2\pi A$ times the force per unit mass-energy, which by use of \eqref{eq:inel} for the inelastic case
and \eqref{eq:force} is given by
\begin{equation}\label{eq:earthen1}
 2 \pi A (m_2/m_1) (v_e/c) \sigma \bar \rho_{\rm s.s.}~~~,
\end{equation}
with $\bar \rho_{\rm s.s.}$
the mean density of sun-bound dark matter along the earth's orbit.  Again writing
$\rho_{m; {\rm s.s.}}=m_2 \rho_{\rm s.s.}$, from \eqref{eq:earthen} and \eqref{eq:earthen1} we get the relation
\begin{equation}\label{eq:sunbound1}
\sigma \bar \rho_{m; \rm{s.s.}} =\frac{1}{4\pi} \frac{dA}{A^2} \frac{v_e}{c} m_1~~~.
\end{equation}
Taking for $dA$ the uncertainty in the change in $A$ over one orbit, this gives
a bound on the product $\sigma \bar \rho_m$ acting over the orbit.

If an inelastic earth-bound dark matter scattering mechanism is responsible for the negative
drag flyby anomalies, then the earth's motion through sun-bound dark matter will
produce acceleration anomalies in the earth's orbit, which (assuming no fine-tuning cancellations) can be used to place
a bound on the density of sun-bound dark matter.  Current bounds
on the yearly change in the radius $A\sim 1.5 \times 10^8 {\rm km}$ of the earth's orbit are around
1.5 m per century \cite {noerd}, or $dA \sim 1.5 {\rm cm}$ per orbit. Substituting
this, the value of $A$, and $v_e \sim 30 {\rm km}{\rm s}^{-1}$ into \eqref{eq:sunbound1} gives the bound\footnote{Since
for an exothermic reaction the cross section varies inversely with velocity for small velocity, the effective cross
section $\sigma$ entering \eqref{eq:sunbound} through \eqref{eq:fest} may be a factor of 3 smaller than that relevant
for the flyby anomaly.}
\begin{equation}\label{eq:sunbound}
\sigma \bar \rho_{m; \rm{s.s.}} \leq 0.5 \times 10^{-31} ({\rm GeV}/c^2){\rm cm}^{-1}~~~.
\end{equation}
However, this formula must be corrected to take into account the fact that for cross sections
$\sigma > 10^{-33} {\rm cm}^2$, the earth diameter exceeds the optical depth for
dark matter scattering on nucleons, and so not all nucleons in the earth have an
equal probability of undergoing a dark matter scattering.  Letting $F_{\rm e}$ denote the
participating fraction of earth nucleons, \eqref{eq:sunbound} must be modified to read
\begin{equation}\label{eq:sunboundmod}
\sigma \bar \rho_{m; \rm{s.s.}}F_{\rm e} \leq 0.5 \times 10^{-31} ({\rm GeV}/c^2){\rm cm}^{-1}~~~.
\end{equation}
In terms of the density of nucleons in earth $\rho_{\rm earth} \sim 3.3 \times
10^{24} {\rm cm}^{-3}$ and the earth diameter $D_{\rm earth} \sim 1.3 \times 10^9 {\rm cm}$, , an estimate of $F$ is
\begin{equation}\label{eq:fest}
F_{\rm e} \sim \frac {1}{\rho_{\rm earth} D_{\rm earth} \sigma} \sim \frac {0.2 \times 10^{-33} {\rm cm}^2 }{\sigma}~~~.
\end{equation}
When substituted into \eqref{eq:sunboundmod}, this gives the bound
\begin{equation}\label{eq:sunboundfinal}
\bar \rho_{m; \rm{s.s.}} \leq 2 \times 10^2 ({\rm GeV}/c^2){\rm cm}^{-3}~~~.
\end{equation}
This bound (which we emphasize depends on the hypothesis of an inelastic dark matter collision explanation for the flyby anomaly, and assumes no cancellation of negative
and positive drag effects for the earth orbit) is considerably lower than the current limit of
$ \sim 10^5 ({\rm GeV}/c^2) {\rm cm}^{-3}$ on excess solar system dark matter.
Taking the ratio of \eqref{eq:ineldens}, which refers to $\bar \rho_{m; {\rm e}}$,
to \eqref{eq:sunboundfinal}, and comparing with \eqref{eq:eacapture}, we learn that
the earth capture fraction in the cascade scenario must obey the constraint
\begin{equation}\label{eq:newratio}
f_e \geq \frac {0.2 \times 10^{-35} {\rm cm}^2 }{\sigma}
~~~.
\end{equation}
For $\sigma =10^{-33}{\rm cm}^2$, this requires the relatively large earth capture
fraction $f_e \geq 0.2 \times 10^{-2}$, but for larger values of $\sigma$ the requirement on $f_e$ becomes less stringent; for example, for $\sigma=10^{-27}
{\rm cm}^2$, \eqref{eq:newratio} becomes $f_e \geq 0.2 \times 10^{-8}$.

We next apply \eqref{eq:sunbound1} to the moon's orbit around the earth.
Lunar
ranging \cite{muller} has established the position of the moon to
within a post-fit residual accuracy of about $2~ {\rm cm}$ relative
to an earth--moon distance of $A_m=384,000 {\rm km}$.  The moon is found to be receding from the earth at a rate of
$3.8 {\rm cm}$ per year, or $0.28 \,{\rm cm}$ per orbit, which is explained by the action of tidal effects.  Estimating the uncertainty in this as $dA_m \sim 0.07 \,{\rm cm}$
for one orbit,  and substituting
this, the moon's orbital velocity $v_m \sim 1 \,{\rm km} \,{\rm s}^{-1}$, and the moon's
orbit radius $A_m$ into \eqref{eq:sunbound1}, we find that along the moon's orbit we must have
\begin{equation}\label{eq:moonbound}
 \sigma \bar \rho_{m; \rm{e}} <\frac{1}{4\pi} \frac{dA_m}{A_m^2} \frac{v_m}{c} m_1
\sim 10^{-29}({\rm GeV}/c^2){\rm cm}^{-1}~~~.
\end{equation}
However, since the radius of the moon is about $0.3$ that of the earth, and the density of the moon is about $0.6$ that of earth,
for cross sections $\sigma > 6 \times 10^{-33}
{\rm cm}^2$, a correction for optical depth $ \sim 6 F_{\rm e} \sim 10^{-33} {\rm cm}^2/\sigma$  is again needed.  Following the
reasoning of \eqref{eq:sunboundmod} through \eqref{eq:sunboundfinal}, we end up with the constraint
\begin{equation}\label{eq:moonboundfinal}
\bar \rho_{m; \rm{e}} \leq   10^4 ({\rm GeV}/c^2){\rm cm}^{-3}~~~.
\end{equation}
Hence the earth-bound dark matter density at the orbit of the moon would have to be many orders of magnitude smaller
than the dark matter density within the radius of 70,000 km relevant for the flyby anomaly.  For example, for an inelastic
cross section $\sigma \sim 10^{-28} {\rm cm}^2$, the dark matter density at the moon's orbit would have to be $10^{-4}$ of
that needed to explain the flyby anomaly, while for an inelastic cross section of $10^{-32} {\rm cm}^2$ it would have to
be a factor $10^{-8}$ smaller.

There are also low altitude constraints
coming from considering satellite orbits.   The satellites of the global
positioning system have orbit radius of $26,600$ km, and geosynchronous satellites
have orbit radii of $\sim 42,000 $ km, but the orbits of these satellites have not been monitored to the level of precision \cite{bender} of that of the LAGEOS geodetic satellite
\cite{rubin}, with orbit radius of $\sim 12,300 $ km.  Residual accelerations of the LAGEOS satellite,  believed to arise from drag effects related to crossings of the earth's
shadow, are smaller in magnitude than $\sim 3 \times 10^{-12} {\rm m} {\rm s}^{-2}$,
as compared with the anomalous flyby accelerations $\sim 10^{-6} 10^4 {\rm m}{\rm s}^{-1}/10^4 {\rm s} = 10^{-6} {\rm m}{\rm s}^{-2}$.  Thus, dark matter densities
at the radius of the LAGEOS orbit would have to be smaller by a factor of  $ 3 \times 10^{-6}$  than at the orbit radii relevant for the flyby anomaly, corresponding to a constraint, in the inelastic case,
\begin{equation}\label{eq:lageosconstr}
\sigma \bar \rho_{m; \rm{e}} \leq 3 \times 10^{-26} ({\rm GeV}/c^2){\rm cm}^{-1}~~~.
\end{equation}
It would clearly be of interest to have comparable anomalous acceleration limits for
the higher-orbiting global positioning system and geosynchronous satellites, since these come closer to the radius 70,000 km relevant for the flyby anomaly.

We consider finally what  the comparable figure would be for Phobos, the moon which orbits Mars
with an orbital radius  of $\sim 9,400 {\rm km}$, with an orbital period $\sim 7 {\rm h} 40 {\rm m}$, an orbital velocity of $\sim 2.1{\rm km}\,{\rm s}^{-1}$,  and an orbital radius decay of $1.8 {\rm cm}\, {\rm y}^{-1}$.  Application of \eqref{eq:moonbound} to
this case, assuming that the residual uncertainty in the orbital decay after taking account of tidal effects is approximately 1/4
of the measured value, gives an upper bound to the dark matter density  at the Phobos orbit
\begin{equation}\label{eq:phobosbound}
\sigma \bar \rho_m < 2 \times 10^{-28} ({\rm GeV}/c^2) {\rm cm}^{-1}~~~,
\end{equation}
which is a factor of 100 tighter than the LAGEOS bound of \eqref{eq:lageosconstr}.
Since $\rho_m$ near Phobos is necessarily greater than the galactic halo density of $0.3 ({\rm GeV}/c^2) {\rm cm}^{-1}$,
this implies that the inelastic cross section is bounded by $\sigma < 7 \times 10^{-28} {\rm cm}^2$.  Conversely, since
we have inferred from the limit on total earth-bound dark matter that $\sigma > 10^{-33} {\rm cm}^2$, we also learn that
near Phobos we must have $\rho_{m; {\rm s.s.}} < 2 \times 10^5 ({\rm GeV}/c^2) {\rm cm}^{-3}$, consistent with known limits on
solar system-bound dark matter.

\subsubsection{Stellar (and solar) dynamics constraints}

Other possible problems  raised by postulating a sun-bound dark matter density larger than
the galactic halo density are whether  the resulting dark matter accretion on
the sun unacceptably alters our well-understood model of solar dynamics, either (i) through additional energy deposition,  (ii) through modifications in energy transport, or (iii) by exceeding the uncertainty in the loss of solar mass from radiation and solar wind.

The possible problem (i) has been discussed in detail, through a running of stellar dynamics codes including dark matter capture, in a recent paper of Fairbairn, Scott and Edsj\"o \cite{scott}, and concludes that ``for a spin-dependent
WIMP-nucleon cross section of $\sigma=10^{-38} {\rm cm}^2$, stars only start to
change their behavior when immersed in a dark matter density of around $10^8$ or
$10^9$ GeV ${\rm cm}^{-3}$'', which corresponds to $\sigma \rho_{m\odot} \sim 10^{-30}~{\rm to} ~10^{-29} ({\rm GeV}/c^2){\rm cm}^{-1}$, with $\rho_{m\odot}$ the dark matter density near the sun.
To convert to a limit on $\bar \rho_{m; {\it s.s.}}$, which we have defined as the density of sun-bound dark matter
near the earth's orbit, we should take account of the fact that the sun-bound dark matter density near the sun may be higher than that near the earth's orbit.  Dividing by a factor of $A/R_{\odot}=1.5 \times 10^8/7\times 10^5=214$, with $R_{\odot}$ the solar radius, as suggested by \eqref{eq:sscapture}, we can write the constraint coming from \cite{scott} as
\begin{equation}\label{eq:scottconstraint}
\bar \rho_{m; {\it s.s.}} \leq \frac {10^{-33} {\rm cm}^2}{\sigma} (5~{\rm to}~50) ({\rm GeV/c^2}){\rm cm}^{-3}~~~,
\end{equation}
which is a highly restrictive limit on the density of sun-bound dark matter.  However, several caveats are needed.  First of
all, since the cross section for an exothermic reaction varies inversely with the velocity, and since the dark matter velocity
near the sun is $\sim 300 {\rm km} \, {\rm s}^{-1}$, which is about 30 times larger than the velocity
$\sim 10 {\rm km} \, {\rm s}^{-1}$ of possible earth-bound
dark matter relevant for the flyby anomaly, the cross section entering \eqref{eq:scottconstraint} is 30 times smaller than the
cross section relevant for the flyby, making the constraint correspondingly less restrictive.  Second, for cross sections larger
than $\sim 10^{-35} {\rm cm}^2$, the optical depth for dark matter colliding with the sun is smaller than the solar radius, and so dark matter with  scattering cross sections from nucleons relevant for the flyby anomaly will not penetrate to the solar core,
changing the way in which dark matter accretion alters solar dynamics.  Finally, the limit of \cite{scott} assumes that dark
matter is self-annihilating, so that the total mass-energy is deposited in the sun.
For non-self-annihilating dark matter, with the secondary dark matter particle weakly interacting so that it escapes from the sun, only the recoiling nucleon kinetic energy is deposited, which is at least 3 orders of magnitude smaller than the primary dark matter particle mass-energy for MeV or lighter dark matter (see \eqref{eq:kin} and \eqref{eq:rat} below).  For these three reasons, we expect the sun-capture constraint of \eqref{eq:scottconstraint} to be substantially weakened, and it should pose no problem \footnote{I wish to thank
Pat Scott for an email pointing out that the stellar dynamics constraint likely requires that we assume the sun-bound dark matter to be non-self-annihilating, and
for bringing the paper \cite{scott} to my attention. It would be interesting to know the limits analogous to those of \cite {scott} for various cases of non-self-annihilating dark matter.   }

The possible problem (ii), dark matter modification of energy
transport in the sun,  has been discussed by Lopes, Bertone, and
Silk \cite{lopes1} and Lopes, Silk, and Hansen \cite{lopes2}, making
use of  constraints coming from helioseismology.  They find that
``in order to be effective in heat transport, WIMPs must have mean
scattering cross section per baryon in the range of $10^{-43} {\rm
cm}^2 \leq \sigma_s \leq 10^{-33} {\rm cm}^2$, depending on the
annihilation cross section and the mass of the WIMP.  The transport
of energy by WIMPs falls rapidly outside of this range.''  The cross
sections of interest for the flyby are mainly larger than the upper
end of this range, and correspond to a dark matter optical depth
that is less than the solar radius, so that dark matter particles
interact with nucleons before penetrating to the solar core.  In the
inelastic, exothermic scenario, dark matter particles in their first
interaction convert to a secondary dark matter particle, that if
weakly interacting escapes from the sun and does not contribute to
energy transport.  So the scenario of \cite{lopes1} and
\cite{lopes2}, in which dark matter particles contribute either to
non-local or localized diffusive energy transport, is not realized,
and does not give constraints.  Additionally, recent revisions in
the chemical composition of the sun \cite{science} appear now to
conflict with current detailed models of the inner dynamics of the
sun, since observed sound speeds are  inconsistent with the values
predicted by solar models using the revised chemical composition. So
it now appears that the helioseismology constraints may not be as
tight as assumed in the papers  \cite{lopes1} and \cite{lopes2}.

 For the
possible problem (iii), one can use the dark matter mass capture rate formula\footnote{This is eq. (11) of \cite{gaisser}, which is a simplification of
(2.31) of Gould \cite{gould}; when the dark matter mass is much larger or smaller
than the nucleon mass, the capture rate is smaller than this estimate.}
\begin{equation}\label{eq:capture}
\dot M \sim \sigma \rho_{m;\odot} (M_{\odot}/m_1) (v_{\rm esc}^2 /v_{\rm dm})~~~,
\end{equation}
with $v_{\rm esc}\sim 620 \,{\rm km} \,{\rm s}^{-1}$ the escape velocity from the sun
and with $v_{\rm dm} \sim 300 \,{\rm km}\, {\rm s}^{-1}$ the velocity of dark matter  near the sun.  Using the bound   $\sigma \rho_{m\odot} \leq 10^{-29} ({\rm GeV}/c^2) {\rm cm}^{-1}$ inferred above
from \cite{scott}, this evaluates to
$\dot M \sim 4 \times 10^{-14}  M_{\odot} {\rm y}^{-1}$, which is smaller
than the estimated  rate \cite{noerd} of solar mass loss from radiation
and solar wind, of $\sim 9 \times 10^{-14} M_{\odot}{\rm y}^{-1}$, and of the same
order as the uncertainties in this rate.  This estimate assumes that the entire mass-energy of the accreted dark matter particle
is retained in the sun; in the inelastic scenario in which the dark matter secondary escapes, the corresponding $\dot M$ will be
much smaller.

\subsubsection{Other astrophysical constraints}

Various astrophysical constraints on dark matter scattering cross sections from nucleons are reviewed in Sec. II A of
Mack, Beacom and Bertone \cite{mack}.  For a dark matter mass $m_2$ smaller than a GeV, these require that the dark matter scattering cross section
from nucleons should be smaller than about $3 \times 10^{-25} (m_2c^2/{\rm GeV}) {\rm cm}^2$, which is compatible with
much of the cross section range inferred from our analysis of the flyby anomaly. (In Sec. II B of \cite{mack}, the authors summarize direct detection constraints, and show that for dark matter masses below a GeV, the entire cross section range
between $10^{33} {\rm cm}^2$ and $10^{-27} {\rm cm}^2$ is allowed.)
Constraints on dark matter interactions
coming from primordial nucleosynthesis have been discussed by Serpico and Raffelt \cite{serp} and by Cyburt, Fields, Pavlidou,
and Wandelt \cite{cyburt}. The former paper gives constraints on the allowed leptonic couplings of MeV range dark matter particles.  The latter studies dark matter scattering from baryons,  when dark matter is also self-interacting, and concludes
that the dark matter scattering cross section from nucleons must be less than about $  10^{-26} (m_2c^2/{\rm GeV}) {\rm cm}^2$, which is more restrictive than the astrophysical constraint of \cite{mack} but is still compatible with much of the
cross section range relevant for the flyby anomaly.

\subsubsection{Earth and satellite heating constraints}

Next, we consider constraints coming from earth and satellite heating, which we shall see are sensitive to the value of the dark matter particle mass $m_2$.  In making
these estimates, we assume that the dark matter is not
self-annihilating; however, the self-annihilating case would give estimates similar to the first case discussed below.  We shall consider bounds
on the dark matter density near the earth's surface
$\rho_{m;R_{\oplus}}$ following \cite{mack} from the earth's heat
flow budget.\footnote{I wish to thank Susan Gardner and John Beacom
for bringing this issue, and reference \cite{mack}, to my attention.}
Let us focus on the inelastic case, and suppose that the cross
section for the primary dark matter particle of mass $m_2$ to
inelastically scatter on a nucleon into the secondary dark matter
particle of mass $m_2'$ is larger than $10^{-33}\, {\rm cm}^2$.  In
this case, the optical depth of the earth is less than one earth
diameter, and an appreciable fraction of earth-intersecting primary
dark matter particles will interact. There are then two limiting
cases to consider.  In the first case, the cross section for
interaction of the secondary dark matter particle of mass $m_2'$ is
also large enough for this particle to be trapped within the earth.
The kinetic energy $\Delta m c^2\sim m_2 c^2$ of this particle is then
dissipated within the earth, and contributes to the earth's heat
flow budget.  The luminosity of the earth is approximately $44 {\rm
TW} \simeq 2.8 \times 10^{23} {\rm GeV}\, {\rm s}^{-1}$, of which
roughly half is accounted for by known mechanisms.  So assuming a
dark matter mass density near earth $\rho_{m;R_{\oplus}}$ with a velocity
of $10^6 {\rm cm}\,{\rm s}^{-1}$, an earth geometric cross section
of $4\pi (R_{\oplus}=6.4 \times 10^8 {\rm cm})^2$, and including
\cite{mack} a solid angle acceptance factor of 1/2, we get the
inequality
\begin{equation}\label{eq:dens1}
\frac{1}{2} \rho_{m;R_{\oplus}} c^2 10^6 {\rm cm}\,{\rm s}^{-1} 4 \pi (6 \times 10^8 {\rm cm})^2
\leq \frac {1}{2} 2.8 \times 10^{23} {\rm GeV}\, {\rm s}^{-1}~~~,
\end{equation}
which gives the restrictive bound\footnote{Since the moon's surface heat flow is less than half that of the Earth,
similar reasoning gives a bound on the dark matter density near the surface of the moon that is half that
of \eqref{eq:dens2}, subject to the caveats that follow.}
\begin{equation}\label{eq:dens2}
\rho_{m;R_{\oplus}} \leq 0.06 ({\rm GeV}/c^2) {\rm cm}^{-3}~~~.
\end{equation}
In the second case, the cross section for interaction of the secondary dark matter
particle is very small, so that it escapes from the earth without interacting.  In this case only the much smaller kinetic energy
\begin{equation}\label{eq:kin}
\delta T_1 \sim m_1 (\delta \vec  v_1)^2/2
\end{equation}
 of the recoiling nucleon is deposited in the earth.  From \eqref{eq:inel}, the ratio of this energy to $\Delta m c^2$ is of order
 \footnote{The DAMA/LIBRA experiment \cite{DAMA} sees an annual modulation signal attributed to 2 to 4 keV nucleon recoils.  If
 we assume $\Delta m \sim m_2$, then \eqref{eq:rat} gives this nucleon recoil energy for a dark matter mass $m_2$ of 2 to 3 MeV.}
\begin{equation}\label{eq:rat}
\frac {\delta T_1}{\Delta m  c^2} \sim \frac{m_2}{2m_1}~~~,
\end{equation}
which gives an effect that depends on the magnitude of $m_2$.  For example, for $m_2$ of order
10 keV, \eqref{eq:rat} is of order $0.5 \times 10^{-5}$, and the bound of \eqref{eq:dens2} is  altered now to
\begin{equation}\label{eq:dens3}
\rho_{m;R_{\oplus}} \leq 10^4 ({\rm GeV}/c^2){\rm cm}^{-3}~~~.
\end{equation}

Although the average effect of dark matter collisions is to alter
the forward velocity of the flyby, there is also a random change in velocity that is
averaged out in \eqref{eq:work}, which would show up as an increase in spacecraft
temperature, as well as in possible localized structural disruption.  Applying  \eqref{eq:kin} in the flyby rest frame  gives an estimate of the thermal energy gain by a nucleon per collision. Dividing this
by the velocity gain by a nucleon per collision
from \eqref{eq:vel}--\eqref{eq:inel}, and multiplying by the total velocity change in the flyby (of order $\sim10^{-6}|\vec u_1|\sim 1 \,{\rm cm}\,{\rm s}^{-1}$), gives the thermal energy gain by a nucleon in the course of the flyby,
\begin{equation}\label{eq:tempgain}
{\rm Temperature ~gain}  \sim
\frac{\delta T_1}{|\delta \vec v_1|} 10^{-6} |\vec u_1|
\sim \frac{1}{2} m_1 |\delta \vec v_1| 10^{-6} |\vec u_1| ~~~.
\end{equation}
In the inelastic case, \eqref{eq:inel} gives $|\delta \vec v_1| \sim m_2 c/m_1$, so that
\eqref{eq:tempgain} becomes
\begin{equation}\label{eq:tempgain1}
{\rm Temperature ~gain}  \sim \frac{1}{2}10^{-6} m_2 |\vec u_1| c
\sim 0.2 {}^{\circ}{\rm K} \Bigg(\frac{m_2 c^2}{ {\rm MeV}} \Bigg)~~~.
\end{equation}
Similarly,  in the elastic case, \eqref{eq:el} gives $|\delta \vec v_1| \sim m_2|\vec u_1-\vec u_2|/m_1$, so that \eqref{eq:tempgain} gives
\begin{equation}\label{eq:tempgain2}
{\rm Temperature ~gain}  \sim \frac{1}{2}10^{-6} m_2 |\vec u_1| |\vec u_1-\vec u_2|
\sim 10^{-5}{}^{\circ}{\rm K} \Bigg(\frac{m_2 c^2}{ {\rm MeV}} \Bigg)~~~.
\end{equation}
These imply that the dark matter mass $m_2$ cannot be too large, or the temperature
gain by the flyby would be noticeable; for example, from the inelastic case
\eqref{eq:tempgain1} we learn that the dark matter mass is constrained to be
substantially less than a GeV.  These results also suggest that sensitive calorimetry in high orbiting spacecraft, and perhaps even sensitive acoustic phonon detection, could be used to
test the hypothesis that the flyby anomalies arise from earth-bound dark matter.

If the dark matter particles are too heavy, collisions with the spacecraft nucleons
will cause recoils energetic enough to produce structural disruption.  If we require that each individual nucleon recoil should not produce structural changes, then we get
a condition of the form
\begin{equation}\label{eq:struc}
\delta T_1 < E_{\rm binding} ~~~,
\end{equation}
with $E_{\rm binding}$ a characteristic atomic binding energy.  In  the inelastic
case, where   \eqref{eq:rat}  with $\Delta m \sim m_2$ gives $\delta T_1 \sim m_2^2 c^2/m_1$, we then get the condition
\begin{equation}\label{eq:struc1}
m_2c^2 < \big(m_1 c^2 E_{\rm binding}\big)^{1/2} \sim 100 {\rm keV}~~~,
\end{equation}
where for sake of
illustration we have taken  $E_{\rm binding}$ as 10 eV.
Again, we see that dark matter particles, if responsible for the flyby anomalies, cannot be too massive.

In a steady-state situation,  the dark matter particle capture at
the earth's surface radius 6,400 km would have to be balanced by
dark matter accumulation from solar system-bound dark matter, at or
above the radius 70,000 km relevant for the flyby anomaly.  Ignoring
evaporation, which should be taken into account in a more careful
estimate, and assuming similar dark matter velocities at radii 6,400
km and 70,000 km,  this gives as the balance condition
\begin{equation}\label{eq:bal}
\rho_{m;s.s.} f_{\rm e} 70^2 \sim \rho_{m;R_{\oplus}} 6.4^2~~~,
\end{equation}
which with $\rho_{m;s.s.} \leq 2 \times 10^2 ({\rm GeV}/c^2)\,{\rm cm}^{-3}$,
and using $f_{\rm e} \leq 1$, gives the
bound
\begin{equation}\label{bal1}
\rho_{m;R_{\oplus}} \leq 2.4\times 10^4  ({\rm GeV}/c^2) \,{\rm
cm}^{-3}~~~.
\end{equation}
This bound, from the steady state condition, is compatible with that of  \eqref{eq:dens3} obtained from the earth heat flow budget.
We conclude that not only must the dark matter density be much smaller near the
moon's orbit than at the radius relevant for the flyby anomaly, but it also must
be substantially depleted near the earth's surface, to be consistent with estimates
based on earth capture.  Whether this depletion would extend to radii as large as
the 30-40 thousand kilometer range relevant for high orbit satellites is not clear.

\subsubsection{Selection rule}

In the inelastic scattering scenario, we are postulating a dark matter primary that scatters from a nucleon into
a lighter dark matter secondary.  If the dark matter primary only interacted with nucleons (or quarks), one could
close the nucleon line into a virtual loop, insert electromagnetic vertices, and deduce a rapid decay of the dark
matter primary into the dark matter secondary plus photons.  To forbid this, the dark matter primary must have interactions
with other particles, which cancel the virtual nucleon contribution to such photon decay loops.  This requires an
appropriate selection rule in the underlying theory of dark and ordinary matter.  For example, consider a symmetry operation
that maps the primary dark matter field into itself, the secondary dark matter field into minus itself, the electromagnetic
field into itself, and interchanges the nucleon (or quark) field with a new baryonic field.  Then the symmetry would
forbid the decay of the dark matter primary into the dark matter secondary plus photons (the loop involving nucleons would
be cancelled by the loop involving the new baryonic field), but scattering of the dark matter primary from nucleons into
the dark matter secondary would be allowed, since the symmetry would only relate this cross section to the corresponding
cross section for scattering of the dark matter primary from the new baryons.

\subsection{Summary}

To summarize, we have made a preliminary survey of whether dark
matter interactions can explain the flyby anomaly.  Our estimates do
not rule out this possibility (for example, we do not find a
requirement that $f_{\rm e}>>1$), but the constraints are severe. To
explain the cases of negative drag flybys, exothermic inelastic
scattering of dark matter on ordinary matter is required.  The cases
of positive drag require either elastic dark matter scattering, or
an asymmetric dark matter velocity distribution in the inelastic
case. In addition, the dark matter must be confined well within the
moon's orbit and depleted near the earth's surface,  a cascade
accumulation mechanism is required to reach the needed dark matter
density, the dark matter mass must be well below a GeV,  the dark
matter interaction cross section with nucleons must be relatively
high (with an inelastic cross section lying between around $10^{-33}
{\rm cm}^2$ and $10^{-27} {\rm cm}^2$), and dark matter must be
non-self-annihilating and stable in the absence of free nucleons.
These constraints can be compatible, since for dark matter masses
much below a GeV, there is little information on nucleon scattering
cross sections,\footnote{See figures 1.7-1.9 of \cite{ann}, which
summarize bounds from experiments searching for multi-GeV dark
matter, and \cite{inel} for light dark matter fits to the DAMA/LIBRA
signal.   For dark matter self-interactions, an analysis
\cite{randall} of the ``bullet cluster'' constrains the ratio of
dark matter self-interaction cross section to mass to be $\sigma/m_2
<0.7 {\rm cm}^2 {\rm g}^{-1}$, which corresponds to $\sigma <
10^{-29}{\rm cm}^2 (m_2 c^2/10 {\rm keV})$. For a related
discussion, see S. L. Adler, arXiv:0808.2823, Phys. Lett. B (in
press). } Further detailed modelling will be needed to see whether
the various constraints can be fulfilled.\footnote{We also remark
that  for dark matter masses close to the electron mass, scattering
from electrons can lead to much higher capture rates than scattering
from nucleons with the same cross section, since the capture rate
formula given in eq. (2.31) of Gould  \cite{gould} scales as the
inverse of the mass $m$ of the particle from which the dark matter
scatters,  at the resonance peak  where the dark matter mass is
equal to $m$. However, the dominant capture mechanism may be
gravitational involving three-body interactions, as simulated in
\cite{xu}.}

One could of course take the severity of the
constraints as an indication that the flyby anomaly must be artifactual, and this
may ultimately turn out to be the case. But if the anomaly is confirmed, and if the
DAMA/LIBRA hints of light dark matter are also confirmed, then
new physics\footnote{Another possibility, corresponding
to ``new physics'',   is
that the anomaly is real and indicates that there is something wrong with our
understanding of electromagnetism \cite{cahill} or gravitation \cite{gravex}.}   will be required, and the scenarios sketched here represent a possibility that merits further exploration.

\subsection{Acknowledgements}

I wish to thank Rita Bernabei for correspondence on the DAMA/LIBRA
experiment, Annika Peter and Edward Witten for  helpful comments,
and  Scott Tremaine for many incisive comments on initial drafts of
this paper.  After the initial arXiv posting of this paper, I
received  email comments from John Beacom, Reginald Cahill,
Jean-Marie Fr\`ere, Susan Gardner, Maxim Yu Khlopov, Annika Peter, and  Pat Scott; many of these comments are reflected in revisions
made in the second posted version. I then spoke about this work at the
conference Quantum to Cosmos - III,  which led to
incisive comments and emails both from the conference organizer Slava Turyshev and  from Peter Bender,  as well as
helpful comments or emails  from Curt Cutler, Viktor Toth, and Myles
Standish.  This further input, as well as emails from  Zurab Silagadze  and Ethan Siegel,  referees' comments, and
comments from Francesco Villante, Francesco Vissante, and other physicists at the Gran Sasso Laboratory, as well as
Alexander Dolgov and other participants in the SpinStat 2008 workshop in Trieste,
led to the current version of this paper.  This work was  supported in part by the Department of Energy under grant no DE-FG02-90ER40542,
and I also benefitted from the
hospitality of the Aspen Center for Physics.


\begin{thebibliography}    {30}


\bibitem{and}
J. D. Anderson, J. K. Campell, J. E. Ekelund, J. Ellis, and J. F.
Jordan, Phys. Rev. Lett. {\bf 100}, 091102 (2008).

\bibitem{DAMA}
R. Bernabei, P. Belli, F. Cappella, R. Cerulli, C. J. Dai, A.
d'Angelo, H. L. He, A. Incicchitti, H. H. Kuang, J. M. Ma, F.
Montecchia, F. Nozzoli, D. Prosperi, X. D. Sheng, and  Z. P. Ye,
``First results from DAMA/LIBRA and the combined results with
DAMA/NaI'', arXiv:astro-ph/0804.2741.

\bibitem{lam} C. L\"ammerzahl, O. Preuss, and H. Dittus, ``Is the
physics within the Solar system really undestood?'', arXiv:gr-qc/
0604052.

\bibitem{inel}
R. Bernabei, P. Belli, F. Cappella, R. Cerulli, C. J. Dai, H. L.
He, A. Incicchitti, H. H. Kuang, J. M. Ma, X. H. Ma, F.
Montecchia, F. Nozzoli, D. Prosperi, X. D. Sheng,  Z. P. Ye, R. G.
Wang, and Y. J. Zhang,  ``Investigation on light dark matter'',
arXiv:astro-ph/0802.4336.

\bibitem{khlop1} M. Yu. Khlopov, ``Cosmoparticle Physics'', World
Scientific, Singapore (1999), Chapters 9 and 10.

\bibitem{khlop2} S. I. Blinnikov and M. Y. Khlopov, { Yad. Fiz.} {\bf 36},
809 (1982), English translation { Sov. J. Nucl. Phys.} {\bf 36}, 472 (1982);
S. I. Blinnikov and M. Yu. Khlopov { Astron. Zh.} {\bf 50}, 632 (1983),
English translation { Sov. Astron.} {\bf 27}, 371 (1983).  For recent discussions
of mirror particles as dark matter, see R. Foot, ``Mirror dark matter and the new
DAMA/LIBRA results: A simple explanation for a beautiful experiment'', arXiv:0804.4518, and Z. K. Silagadze, ``Mirror dark matter discovered?'',
arXiv:0808.2595.

\bibitem{wein1}  S. Weinberg, {\it The Quantum Theory of Fields I}, pp. 156-157
(Cambridge University Press, 1995).

\bibitem{dark}  S. L. Adler, ``Placing direct limits on the mass of earth-bound
dark matter'', arXiv:0808.0899.

\bibitem{kr}
T. Damour and L. M. Krauss, Phys. Rev. D {\bf 59}, 063509 (1999).

\bibitem{ann}
A. H. G. Peter, ``Particle Dark Matter in the Solar System'',
unpublished Princeton University dissertation, to be submitted
June 2008.

\bibitem{xu} X. Xu and E. R. Siegel, ``Dark Matter in the Solar System'',
arXiv:0806.3767.

\bibitem{frere} J.-M. Fr\`ere, F.-S. Ling, and G. Vertongen,
{ Phys. Rev. D} {\bf 77}, 083005 (2008).

\bibitem{khrip}
I. B.Khriplovich and E. V. Pitjeva, { Int. J. Mod. Phys. D}
{\bf 15}, 615 (2006); M. Sereno and Ph. Jetzer, Mon. Not. R. Astron.
Soc. {\bf 371}, 626 (2006); L. Iorio, JCAP {\bf 0605}, 002 (2006); I. B. Khriplovich, { Int. J. Mod. Phys. D}{\bf 16}, 1475 (2007).

\bibitem{annmod}
A. K.  Drukier, K. Freese, and D N. Spergel,  Phys. Rev. D {\bf 33}, 3495 (1986); K. Freese,
J. Frieman, and A. Gould,  Phys. Rev. D {\bf 37}, 3388 (1988).

\bibitem{noerd} P. D. Noerdlinger, ``Solar Mass Loss, the Astronomical Unit, and
the Scale of the Solar System'', arXiv:0801.3807, which quotes Pitjeva, 2008 (private communication) as giving $\dot A \sim 1.5 {\rm m}{\rm cy}^{-1}$.

\bibitem{muller} J. M\"uller, J. G. Williams, and S. G. Turyshev,
``Lunar Ranging Contributions to Relativity and Geodesy'',
arXiv:gr-qc/0509114.

\bibitem{bender}  I wish to thank Peter Bender (private communication)  for emphasizing the importance
of the constraints coming from LAGEOS.

\bibitem{rubin} D. P. Rubincam, J. Geophys. Res. {\bf 95}, 4881 (1990);
R. Sharroo, K. F. Wakker, B. A. C. Ambrosius, and R. Noomen, J. Geophys. Res..
{\bf 96}, 729 (1991).

\bibitem{scott} M. Fairbairn,
P. Scott, and J. Edsj\"o, ``The zero age main sequence of WIMP burners'',
arXiv:0710.3396; see also  P. Scott, J. Edsj\"o, and M. Fairbairn, ``Low mass stellar
evolution with WIMP capture and annihilation'', arXiv:0711.0991.

\bibitem{lopes1}  I. P. Lopes, G. Bertone, and J. Silk, Mon. Not. R. Astron. Soc. {\bf 337},
1179 (2002).


\bibitem{lopes2}  I. P. Lopes, J. Silk, and S. H. Hansen, Mon. Not. R. Astron. Soc.
{\bf 331}, 361 (2002).

\bibitem{science} M. Asplund, Science {\bf 322}, 51 (2008).


\bibitem{gaisser}  T. K. Gaisser, G. Steigman, and S. Tilav, Phys. Rev. D {\bf 34},
2206 (1986).


\bibitem{gould}  A. Gould, Astrophys. J. {\bf 321}, 571 (1987).


\bibitem{mack} G. D. Mack, J. F. Beacom, and G. Bertone,  Phys. Rev. D
{\bf 76}, 043523 (2007).

\bibitem{serp} P. D. Serpico and G. G. Raffelt, Phys. Rev. D {\bf 70}, 043526 (2004).

\bibitem{cyburt} R. H. Cyburt, B. D. Fields, V. Pavlidou, and B. D. Wandelt, Phys. Rev.
D {\bf 65}, 123503 (2002).



\bibitem{randall} S. W. Randall, M. Markevitch, D. Clowe, A. H. Gonzalez, and
M. Brada\v c, Astrophys. J. {\bf 679}, 1173 (2008).

\bibitem{cahill} R. T. Cahill, Prog. Phys. {\bf 3}, 9 (2008);  W. Petry, ``A Possible Explanation of Anomalous Earth Flybys'', arXiv:0806.0334.

\bibitem{gravex}  H. J. Busack, ``Simulation of the flyby anomaly by means of an
empirical asymmetric gravitational field with definite spatial orientation'',
arXiv: 0711.2781; A. Unzicker, ``Why do we Still Believe in Newton's Law? Facts,
Myths and Methods in Gravitational Physics'', arXiv:gr-qc/0702009;
M. E. McCulloch, ``Modelling the flyby anomalies using a
modification of inertia'', arXiv:0806.4159; M. B. Gerrard and T. J. Sumner,
``Earth Flyby and Pioneer Anomalies'', arXiv:0807.3158.

\end{thebibliography}
\end{document}